\let\myover=\over   
\documentclass[12pt,a4paper]{article}
\usepackage{epsfig}
\def\be{\begin{equation}}
\def\ee{\end{equation}}
\def\d{\partial}

\def\be{\begin{equation}}
\def\ee{\end{equation}}
\def\d{\partial}
\def\half{\frac{1}{2}}

\newcommand{\bg}{\begin{gather}}
\newcommand{\eg}{\end{gather}}

\begin{document}
\let\over=\myover  
\def\half{{1 \over 2}}

\begin{titlepage}
\thispagestyle{empty}

\vspace*{2cm}
\begin{center}
\Large \bf Simplification of Flavour Combinatorics\\
        in the Evaluation of Hadronic Processes 
\end{center}

\vspace*{1.5cm}

\begin{center}
\large 
E.E.~Boos$^*$, V.A.~Ilyin$^*$  and  A.N.~Skachkova$^{*\dagger}$
\end{center}

\begin{center}
{\small\em
$^*$Institute of Nuclear Physics, Moscow State University, 119899
Moscow,
Russia} \\
{\small\em
$^\dagger$Joint Institute for Nuclear Research, 141980 Dubna, Russia
}       
\end{center} 
       
\begin{quotation}
\vspace*{1cm}
        
\vskip 2cm  
\begin{center}
{\sf Abstract}
\end{center}
{\small 
\vspace*{0.3cm}

A serious computational problem in the evaluation of hadronic collision
processes is connected with the large number of partonic subprocesses
included in the calculation. These are from the quark and gluon content of
the initial hadrons, and from CKM quark mixing. For example, there are 180
subprocesses which contribute to the $W$+2jets process, and 292 
subprocesses in $W$+3jets production at the LHC, even when quarks from 
only the first two generations are taken into account.
 
We propose a simple modification of the rules for evaluation of cross
sections and distributions, which avoids multiplication of channels from
the mixture of quark states. The method is based on a unitary rotation
of down quarks, thus, transporting the mixing matrix elements from
vertices
of Feynman diagrams to the parton distribution functions (PDF). As a
result, one can calculate cross sections with significantly fewer
subprocesses. For the example mentioned above, with the new rules, one
need evaluate only 21 and 33 subprocesses respectively. The matrix
elements of the subprocesses are calculated without quark mixing but with 
a modified PDF convolution which depends on the quark mixing angle, and
on the topologies of gauge invariant classes of diagrams. The proposed
method has been incorporated into the CompHEP program and checked with
various examples.

} 
\end{quotation}

\end{titlepage}

\vspace{2cm}
\noindent
{\large \bf Introduction}
\vspace{0.2cm}

The evaluation of cross sections for hadron collisions typically involves a
large number of subprocesses. The reason for this is the quark and gluon
content of the initial hadrons and the Cabbibo-Kobayashi-Maskava (CKM)
mixing of the down quarks. At Tevatron and LHC energies in many cases 5
flavors, $u$, $d$, $c$, $s$ and $b$, give a sizable contribution through
the corresponding parton densities. Note that subprocesses sometimes
contribute only because of the non-diagonality of the CKM matrix. An
example is the subprocess $u\bar d\to s\bar d W^+$ which contributes to
the $W+2jets$ production. Furthermore, the number of subprocesses increases
with the number of quarks in the final state, in particular, because the 
final state jets produced by various subprocesses are 
indistinguishable \footnote{An
important exclusion is a $b$-quark jet which can be distinguished  from light
quark jets with a micro-vertex detector.}. The process of $W$-boson and
jet production (an important background to various Standard Model and "new
physics" processes) exemplifies the problem: 180 subprocesses contribute to
$W+2jets$ and 292 subprocesses to $W+3jets$ production if only the  quarks
of first two generations and only QCD diagrams are taken into account.

The huge multiplication of channels stands as a real computation problem
despite the fact that the matrix elements of some subprocesses may have
similar analytical structures. Indeed, one should separately organize Monte
Carlo  integration and/or event generation for each subprocess separately
because the convolution with the parton distribution functions (PDF) is
flavour dependent. This is a present-day problem for automatic calculations
of collision processes (see e.g. \cite{reviews} and references therein for
review of this new computation approach).

In this letter we propose a new method which simplifies the flavour
combinatorics and reduces the multiplication of channels due to the mixture
of quark states. Implementation of the proposed algorithm in Monte Carlo
codes is straightforward. It has been incorporated into the CompHEP program
\cite{CompHEP} and checked for many examples. In the method the quarks of
the first two generations are taken to be massless and do not mix with the
third generation  which is obviously a good approximation for many
processes in the Tevatron and LHC energy range. The method is based on a
rotation of down quarks, thus, transporting the mixing matrix elements from
vertices of subprocess Feynman diagrams to the parton distribution
functions. The complete set of subprocess diagrams are divided into gauge
invariant classes, and depending on the topology of the class, new
computation rules are formulated. The method drastically simplifies the
calculation of matrix elements but leads to a modification of the PDF
convolution procedure. We demonstrate the power of the proposed technique
with the example of $W+2jets$ production at LHC. Here only 21 subprocesses
have to be evaluated with the new technique, which should be compared with
180 subprocesses in the standard approach.

The paper is organized as follows. In Section 1 we discuss CKM
diagonalization at the level of Feynman amplitudes and introduce basic
notations. Here the topologies of the gauge invariant classes of diagrams are
defined, for which in Section 2 new PDF convolution rules are formulated.
The $W+2jets$ production process at LHC is discussed in Section 3. In  the
Appendix we briefly comment on the implementation of the new rules in the
CompHEP program.  

\vspace{0.5cm}
\noindent
{\large \bf 1. Feynman amplitudes}
\vspace{0.2cm}

\begin{figure}[hb]
\unitlength=1cm
\vspace*{1cm}
\begin{picture}(16,14)
\put(0,13.5){\mbox{a)}}
\put(2,10){\epsfxsize=4.5cm \leavevmode \epsfbox{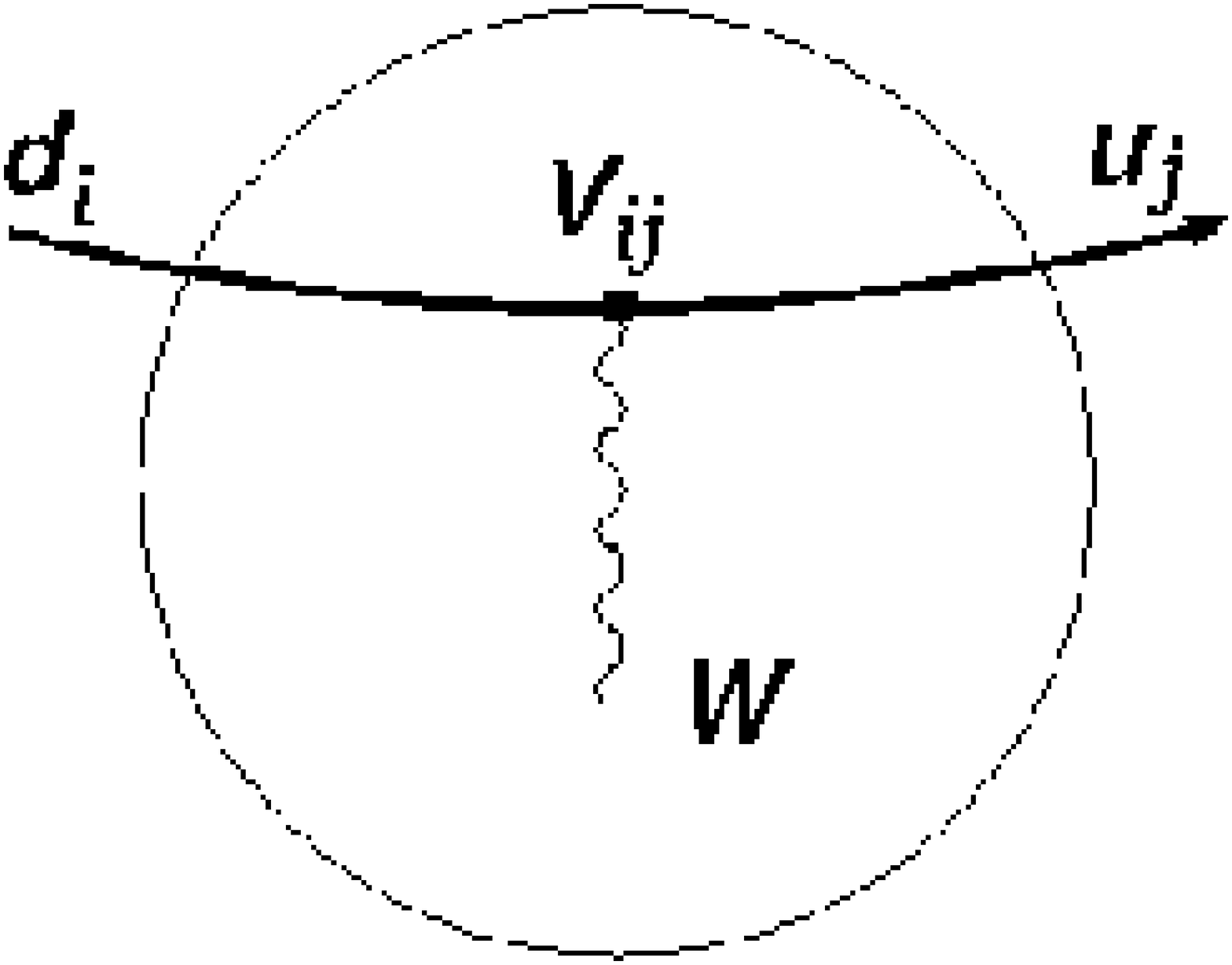}}
\put(7.5,11.5){\mbox{\LARGE $\Longrightarrow$}}
\put(9.5,11.5){\mbox{\LARGE $V_{ij}\;\cdot$}}
\put(11,10){\epsfxsize=4.5cm \leavevmode \epsfbox{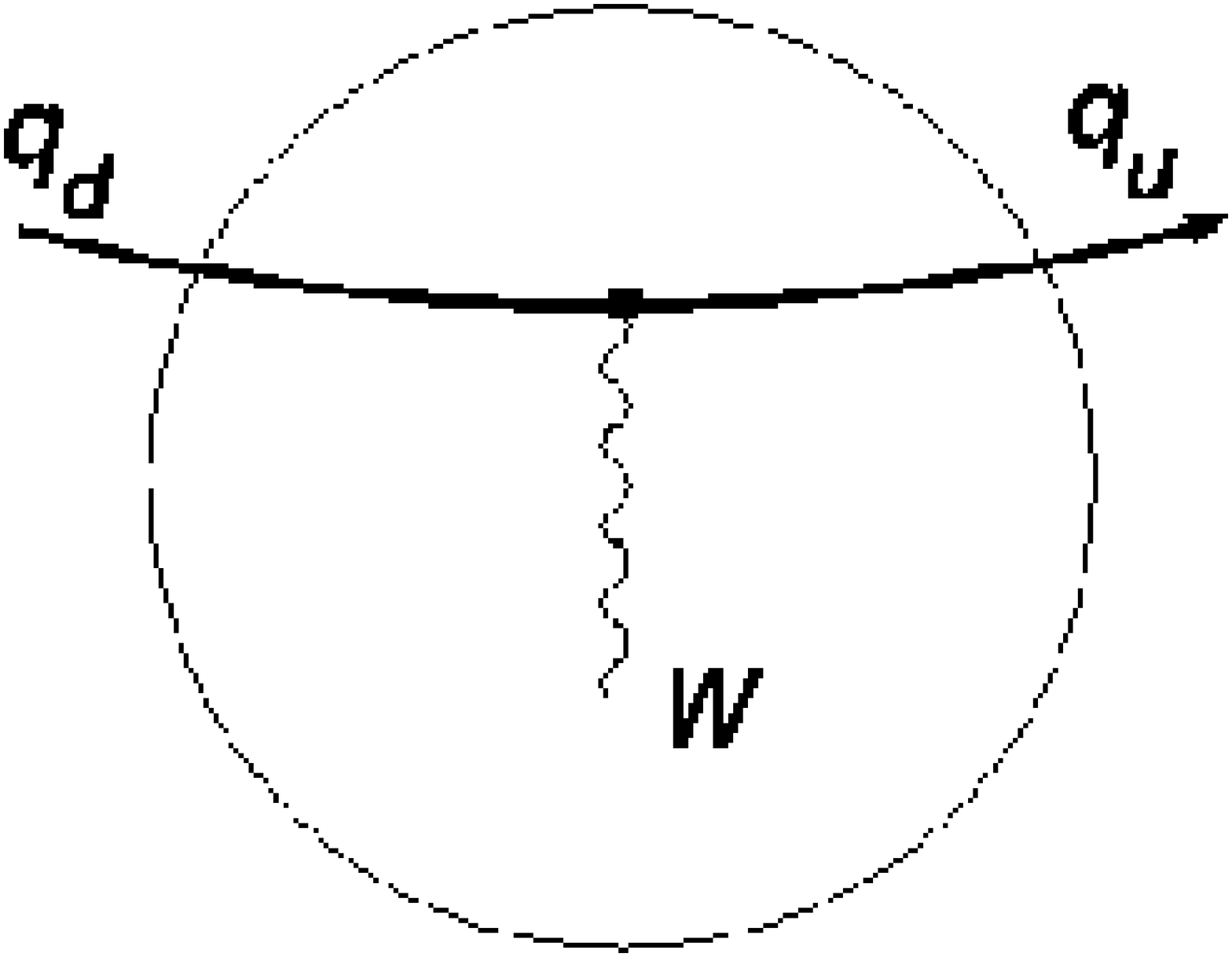}}
\put(0,8.5){\mbox{b)}}
\put(1,6.5){\mbox{\LARGE $\Sigma_k$}}
\put(2,5){\epsfxsize=4.5cm \leavevmode \epsfbox{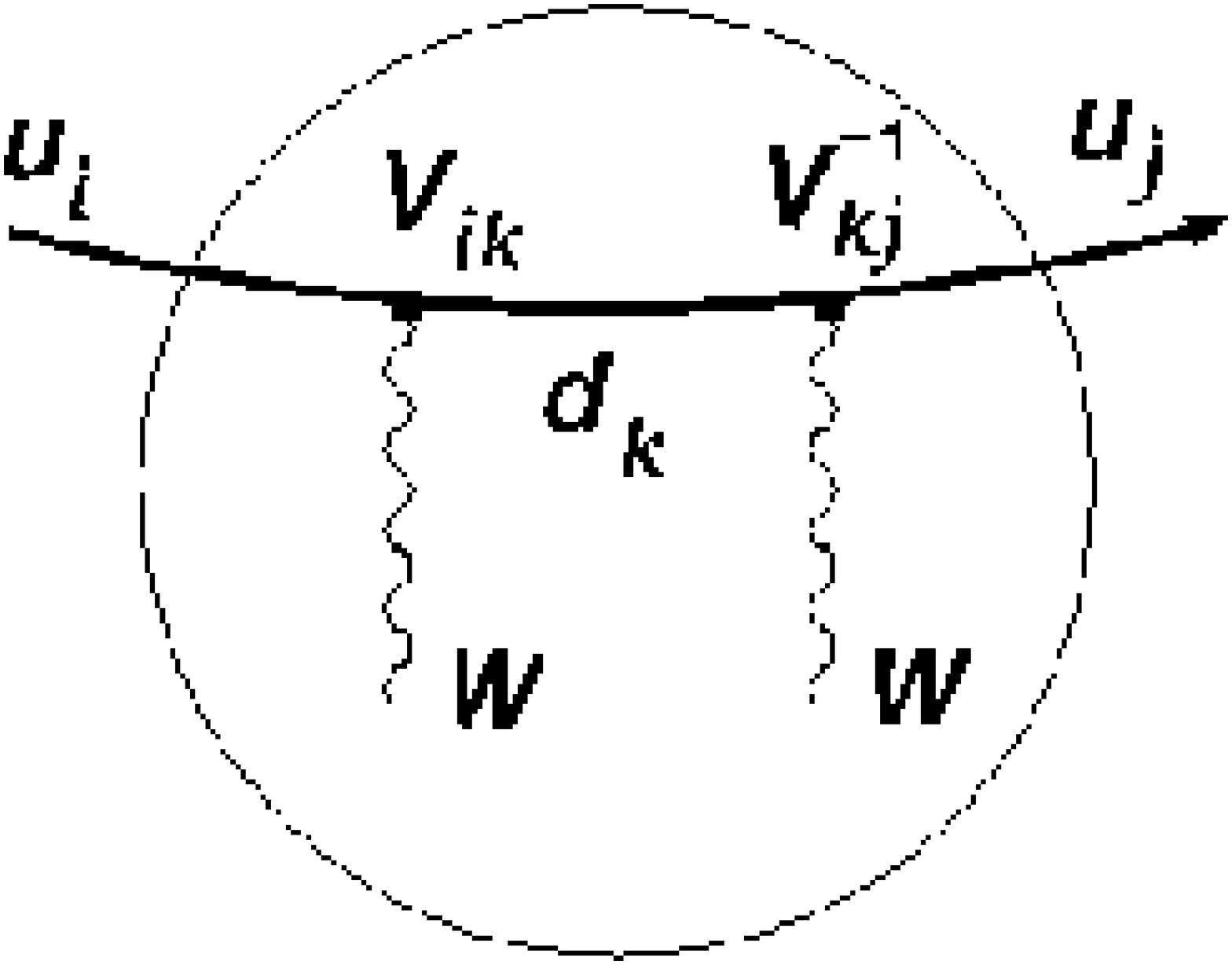}}
\put(7.5,6.5){\mbox{\LARGE $\Longrightarrow$}}
\put(9.5,6.5){\mbox{\LARGE $\delta_{ij}\;\cdot$}}
\put(11,5){\epsfxsize=4.5cm \leavevmode \epsfbox{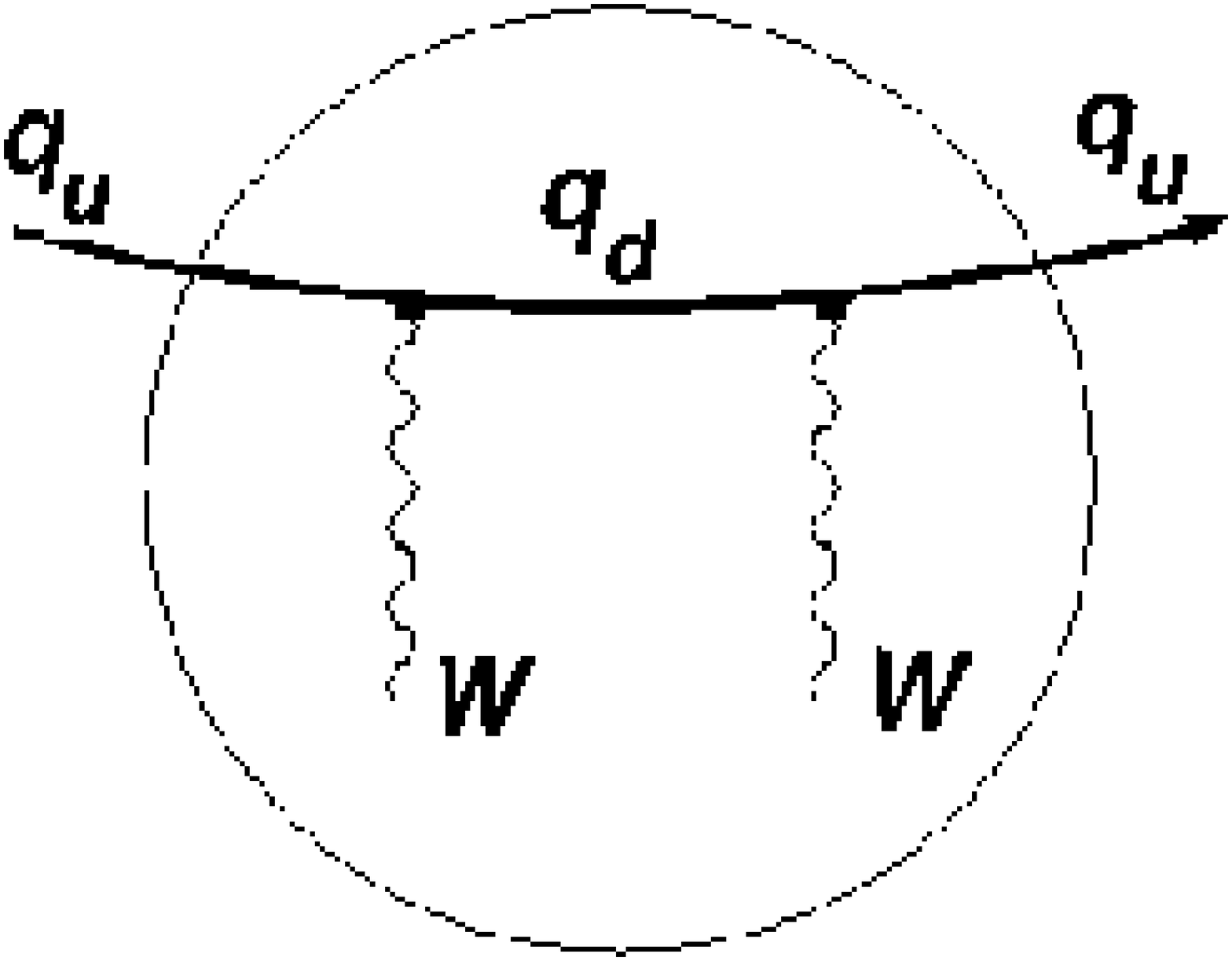}}
\put(1,2){\mbox{\LARGE $\Sigma_k$}}
\put(2,0.5){\epsfxsize=4.5cm \leavevmode \epsfbox{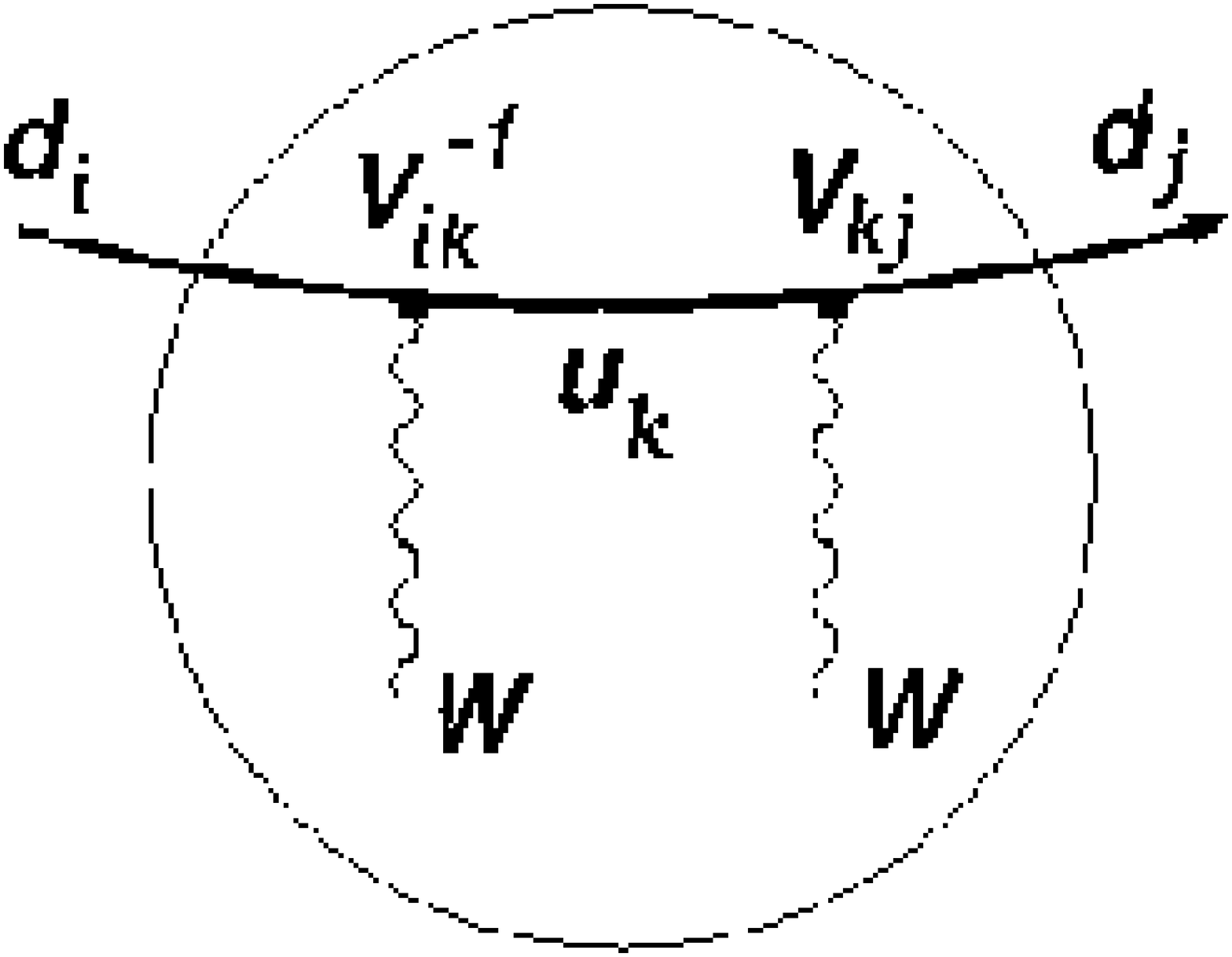}}
\put(7.5,2){\mbox{\LARGE $\Longrightarrow$}}
\put(9.5,2){\mbox{\LARGE $\delta_{ij}\;\cdot$}}
\put(11,0.5){\epsfxsize=4.5cm \leavevmode \epsfbox{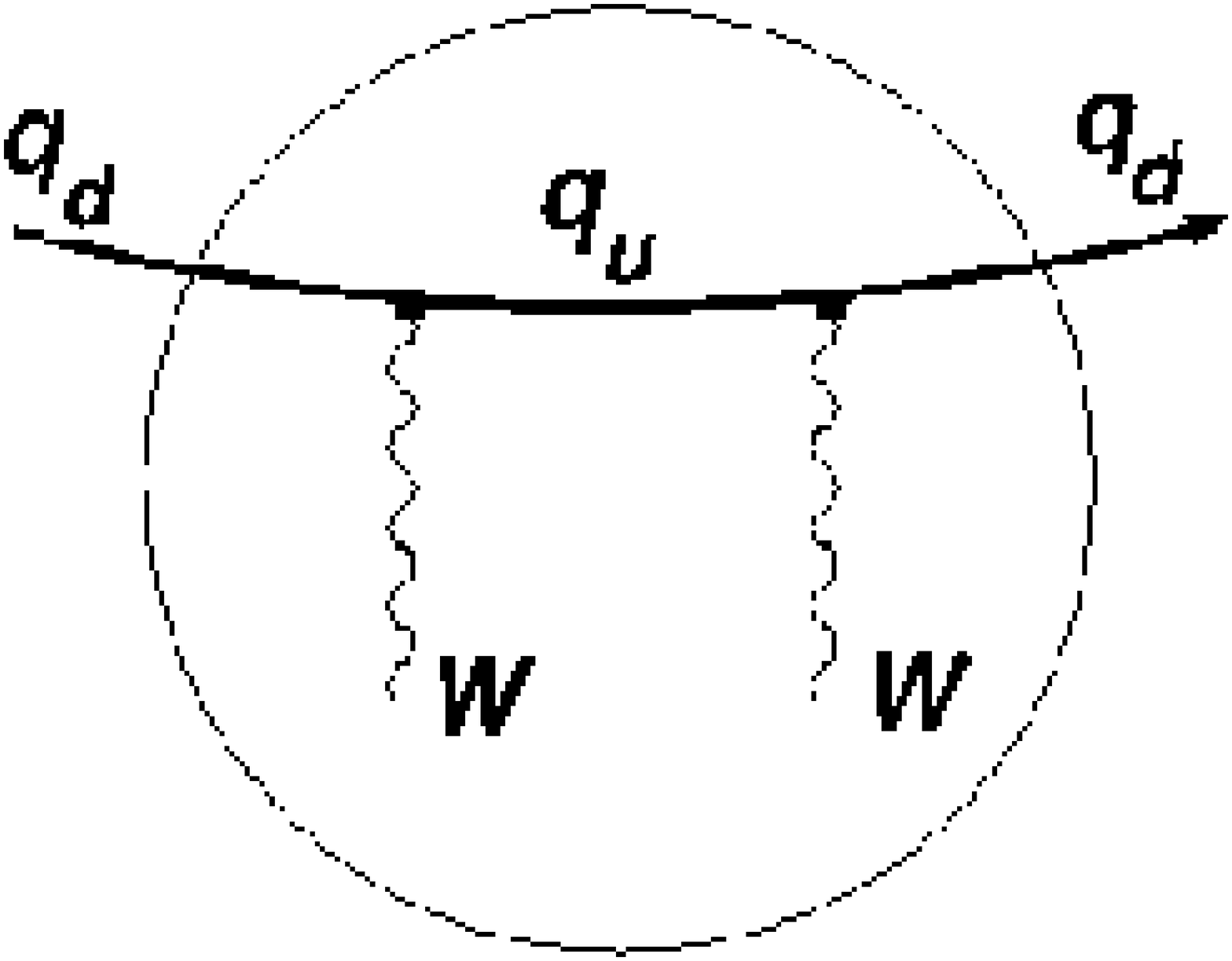}}
\end{picture}
\caption{Generic scattering topology diagrams with:
a) a single $W$ vertex  and b) two $W$ vertices on the quark
line. No summation over the generation indices $i$ and $j$ is implied. 
Here and in the next figure the external lines on the left and right
sides of the diagram correspond
to {\it in}- and {\it out}-states respectively.
\label{fig1} }
\end{figure}

\begin{figure}[hb]
\unitlength=1cm
\begin{picture}(16,7.0)
\put(1,7.5){\mbox{a)}}
\put(2,5){\epsfxsize=5.0cm \leavevmode \epsfbox{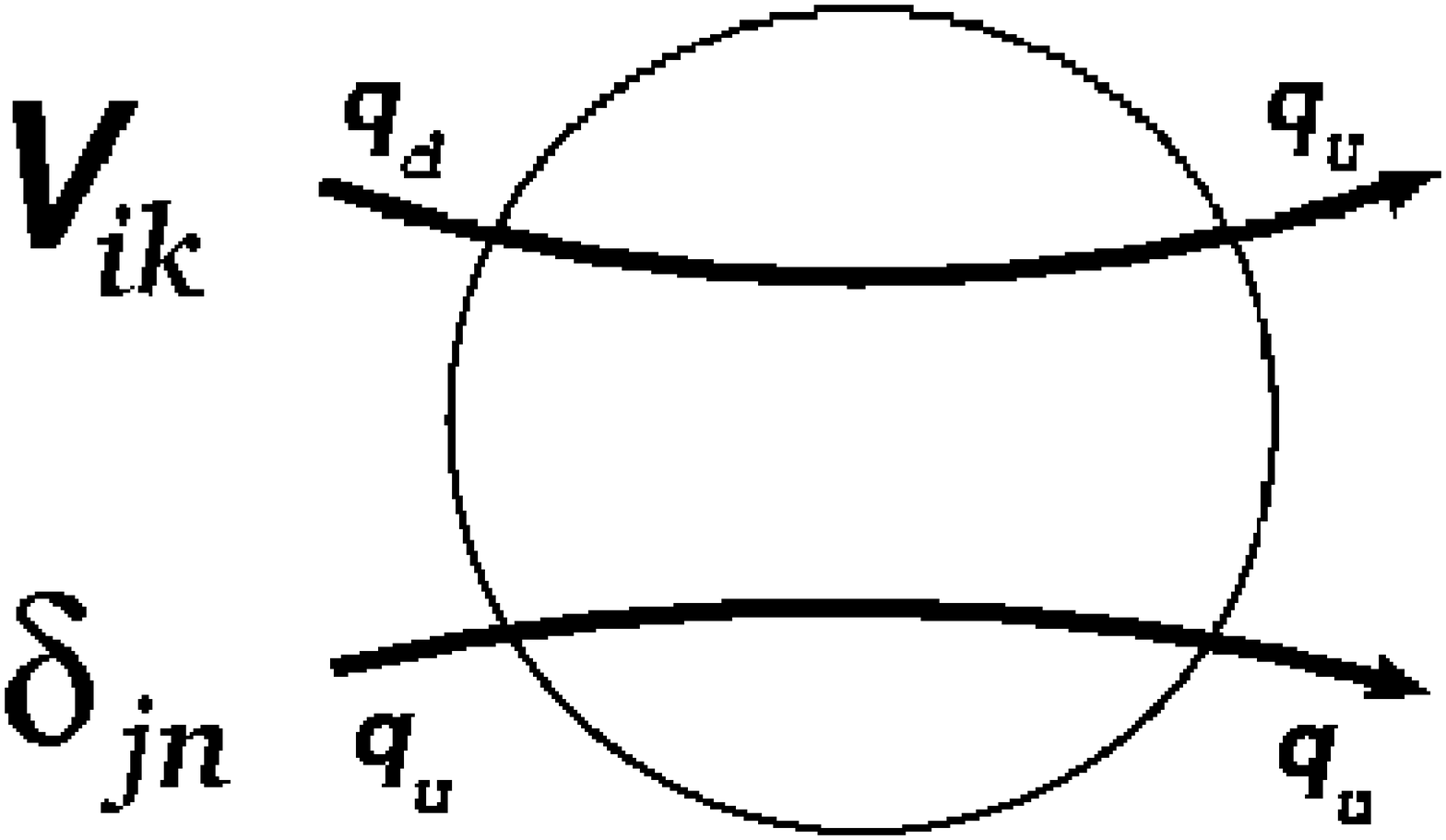}}
\put(10,7.5){\mbox{b)}}
\put(11,4.7){\epsfxsize=4.5cm \leavevmode \epsfbox{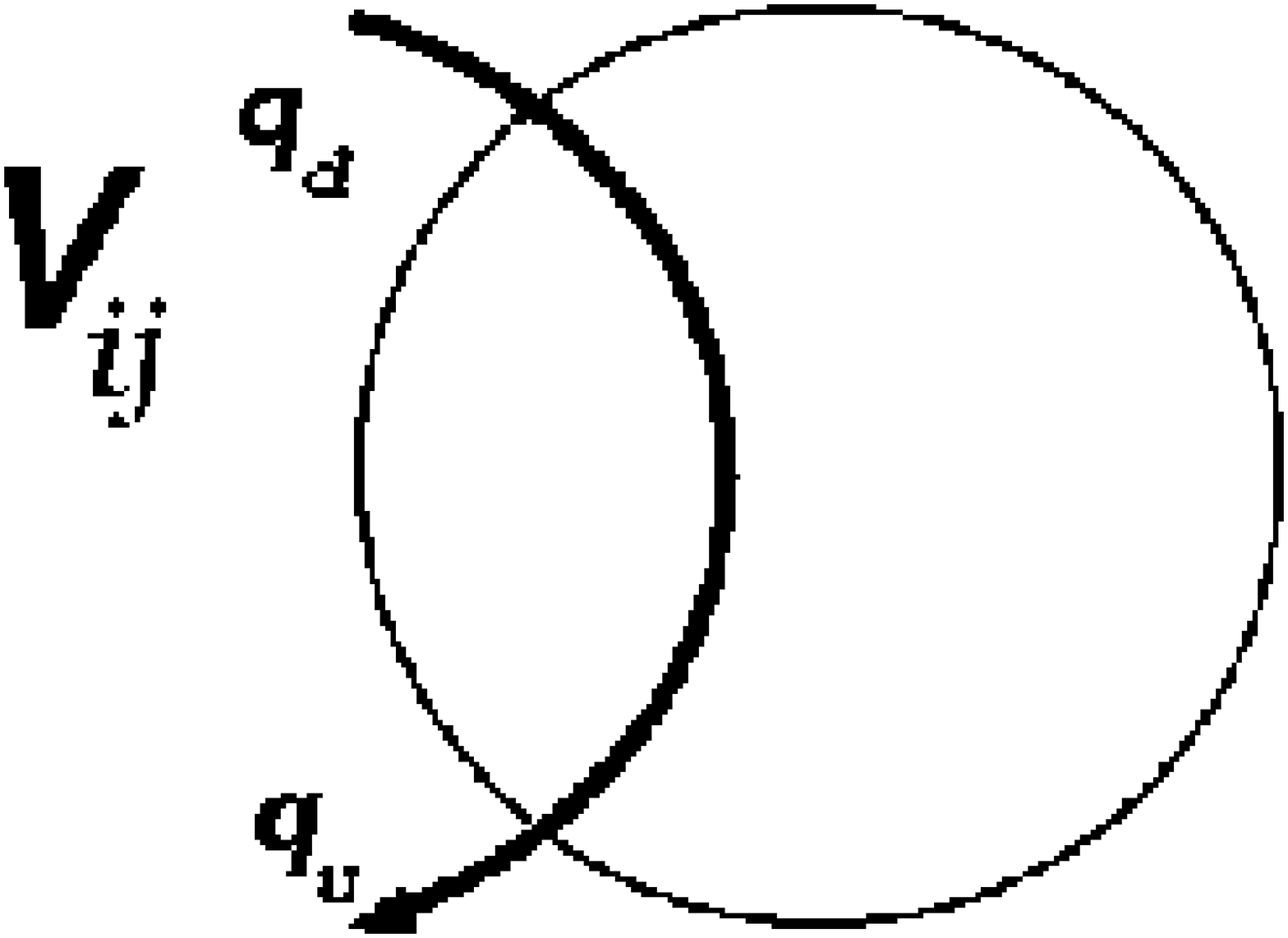}}

\put(1,3.5){\mbox{c)}}
\put(2,0.5){\epsfxsize=4.5cm \leavevmode \epsfbox{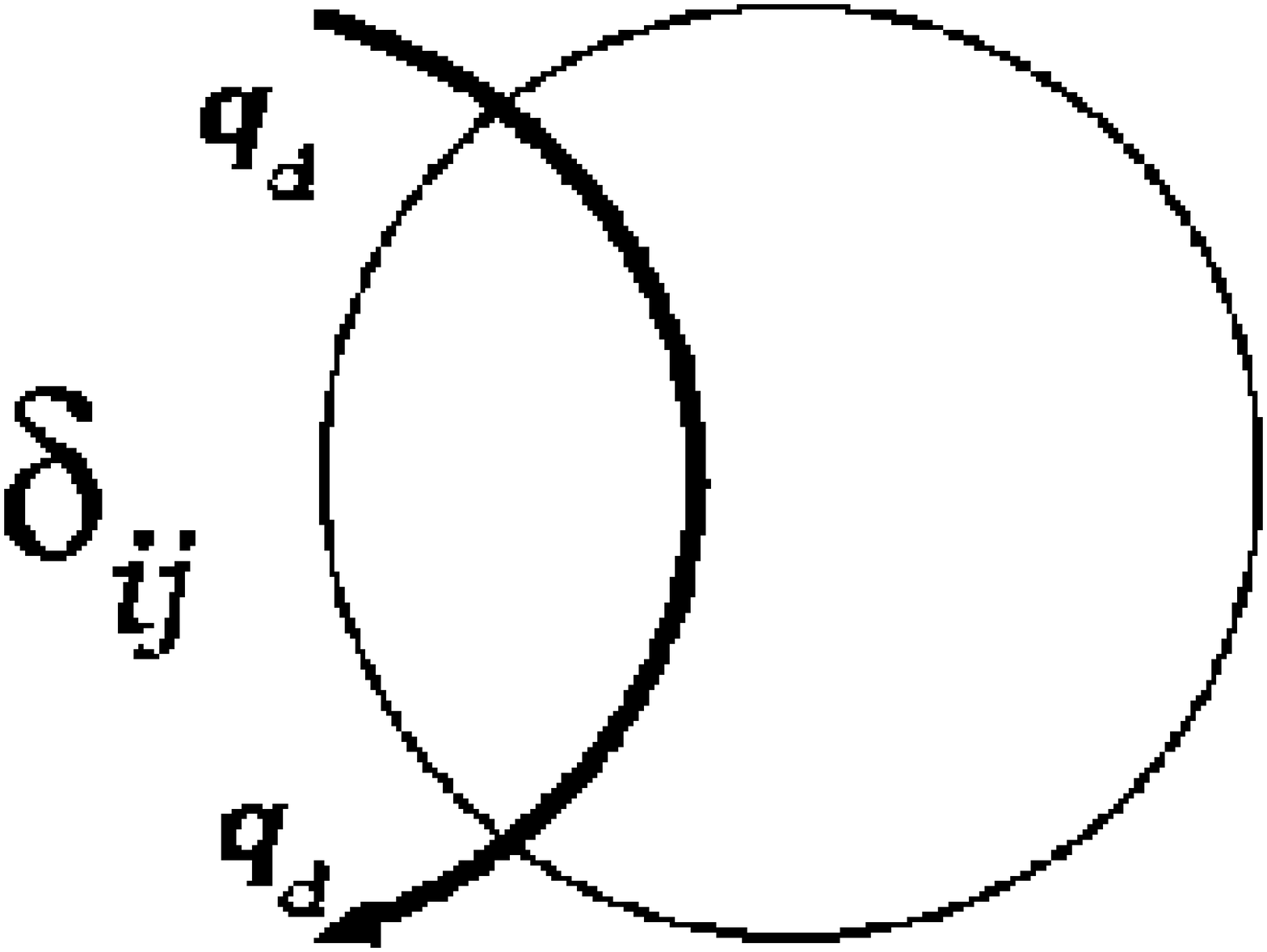}}
\put(10,3.5){\mbox{d)}}
\put(11,0.5){\epsfxsize=4.5cm \leavevmode \epsfbox{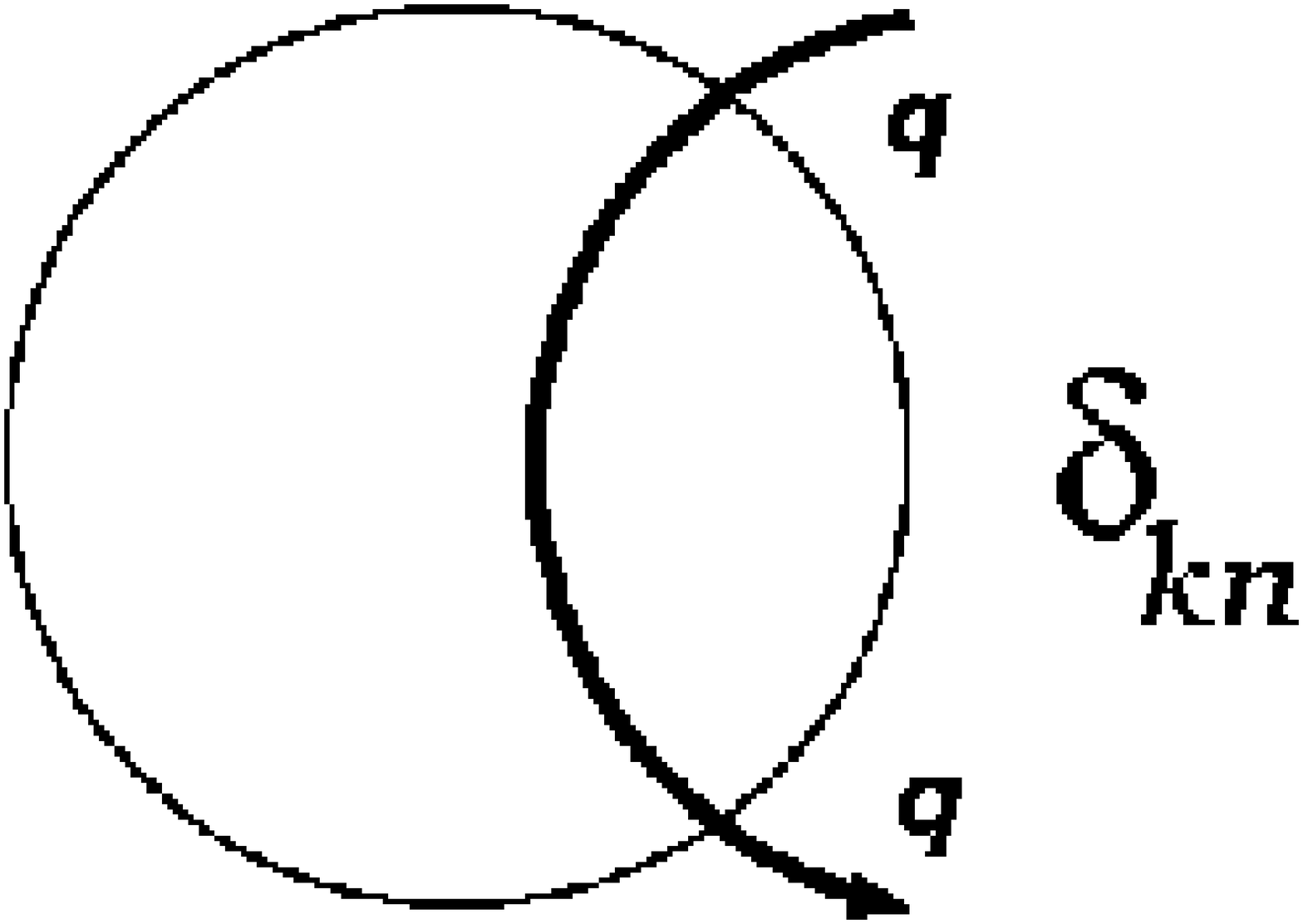}}
\end{picture}
\caption{Generic diagrams with different topologies 
in which the quark lines connect
{\it in}- and {\it out}-states. Here $i$ and $j$ are
the generation indices of the {\it in}-quarks, while $k$ and $n$ 
are the generation indices for
{\it out}-quarks. Diagram d) includes a quark line connecting two {\it
out}-states.
\label{fig2} }
\end{figure}

\begin{figure}[hb]
\unitlength=1cm
\begin{center}
\begin{picture}(16,4.0)
\put(0.6,4.3){\mbox{a)}}
\put(0.6,0.3){\mbox{b)}}
\put(-1,-8.5){\epsfxsize=10cm \leavevmode \epsfbox{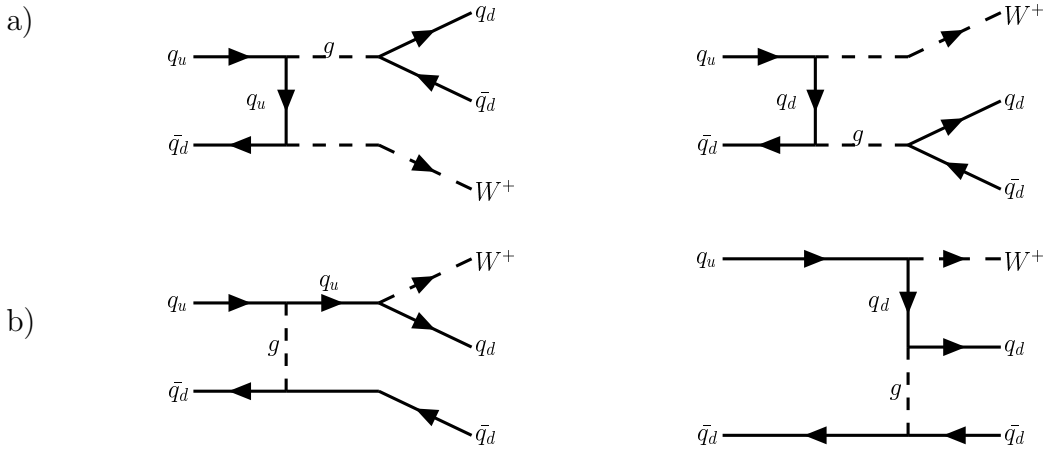}}
\end{picture}
\end{center}
\vspace*{1.0cm}
\caption{Feynman diagrams for the subprocess $q_u \bar{q_d}\to q_d
\bar{q_d} W^+$ which
contribute to $W^+ +2jets$ production at LHC.
Case a) corresponds to the ${\cal D}_{a}^{CC}$ topology of diagrams,
while case b) corresponds to the ${\cal D}_{sc}$ topology.
\label{fig3} }
\end{figure}

Let us consider a parton subprocess with a quark in the initial state. 
There are two possible topologies of any Feynman diagram with  the
corresponding quark line: 1) the quark line goes through the diagram from
the initial to final state -- "scattering topology", and 2) the quark line
connects both initial state partons -- "annihilation topology". In Fig.\ref{fig1}
and Fig.\ref{fig2}a one can see examples of scattering topology, and
diagrams of annihilation topology are shown in Fig.\ref{fig2}b,c. Here the
interior of the circle represents the diagram, and the solid line
represents a quark line (quark current). An important point is that
according to the theorem proved in \cite{BO}, the two sets of diagrams with
the scattering and annihilation topologies form gauge invariant classes
with respect to the SM gauge group. Thus, one can change the computational
rules for each class independently. Note, that a specific parton
subprocess could contain only scattering topology diagrams, or only
annihilation topology diagrams, or both. For the process $u \bar{d}\to d
\bar{d} W^+$ shown in Fig.\ref{fig3} both annihilation and scattering
diagrams contribute. For our further consideration it is necessary  to
determine whether the quark current is charged (CC) or neutral (NC). This is
done by counting the quark vertices with $W$-bosons. The charged current
involves an odd and the neutral current an even number of $W$ vertices.
Both cases might include an arbitrary number of vertices with gluons,
photons and Z-bosons.  

Two approximations are used in our method. We neglect: 1)the masses of
quarks from the two first generations, and  2) the mixing with the third
 generation.
Thus, the $3\times 3$ CKM mixing matrix reduces to the $2\times 2$ Cabbibo matrix:
$$ V_{CKM}  \;\Longrightarrow\; \left( \begin{array}{cc}
                                            V & 0 \\
					    0  & 1    
				       \end{array} \right)\;,
   \qquad
   V \;=\;  \left( \begin{array}{cc}
                         \cos\vartheta_c  & \sin\vartheta_c \\
     	                 -\sin\vartheta_c & \cos\vartheta_c   
				       \end{array} \right)$$
where $\vartheta_c$ is the Cabbibo angle. 

The zero quark mass limit is a good approximation in many applications, in
particular, when one evaluates matrix elements of hard subprocesses.
Indeed, the energy scale, e.g. the partonic collision energy, for 
Tevatron-LHC hard subprocesses is ${\cal O}(100)$ GeV and higher. Then,
minimal energy-like cuts on jets (momentum transfer, invariant masses of
jet pairs etc.) are of order ${\cal O}(10)$ GeV or more. This means that the
momenta in the propagators are of the same order and mass effects can be
neglected in matrix elements. At the same time, the flavour dependence of
parton distributions is sizable and can substantially affect observables.

The second approximation (no mixing with the third generation) also works
well in many applications. The $V_{tb}$ mixing matrix element is very close
to unity and the non-diagonal CKM matrix elements for bottom and top quarks
are very small. Of course, these elements are responsible for important
physical phenomena, such as B-meson oscillations and rare decays of heavy
mesons. In these cases the proposed method can, obviously, not be applied.
Another reason why the second approximation is reasonable is that $t$ and
$b$ quarks produce final state objects in a detector which are very
different from light quark jets: the $b$ quark jet has a secondary vertex,
and the $t$-quark appears as a heavy narrow resonance.

The starting point of our consideration is the diagonalization of the quark
mixing matrix in vertices. This means that rotated down quark states are used
in Feynman rules, $d'=d\cos{\vartheta_c}+s\sin{\vartheta_c}$ and
$s'=-d\sin{\vartheta_c}+s\cos{\vartheta_c}$ rather than $d$ and $s$ states
being eigenstates of the mass matrix. It is worth recalling that all
electroweak vertices are diagonal over the isodoublets {\scriptsize
$\left(\begin{array}{c} u\\d'\end{array}\right)_L$ } and {\scriptsize
$\left(\begin{array}{c} c\\s'\end{array}\right)_L$ }. For example, the
charged current (CC) electroweak vertex is diagonal in this rotated basis,
$$
 W^+_\mu \cdot J^\mu_{CC} \;=\; W^+_\mu \cdot 
   {\bar{u}_i}^L \,\gamma^\mu  \,V_{ij} {d_j}^L \;=\;
   W^+_\mu \cdot 
   {\bar{u}_i}^L \,\gamma^\mu  \,{d'_j}^L\;.
$$
Here $ d'_i = V_{ij} d_j$ (where $i,j=1,2$ are the generation indices, so
$d=d_1$ and $s=d_2$). As a result, elements of the CKM mixing matrix do not
enter in the matrix element if one calculates in terms of these rotated down
quarks.

Let us consider an electroweak model with the only one generation of quarks
(referred to as the {\it $EW_{ud}$} model), and denote generalized up and down
quarks by $q_u$ and $q_d$ respectively. Then, Feynman amplitudes in the EW
model with two generations and Cabbibo mixing can be evaluated in the
$EW_{ud}$ model with a single quark generation. Indeed, in the $EW_{ud}$ model,
the CC vertex represents the Lagrangian term $W^+_\mu{\bar{q_u}}^L
\gamma^\mu {q_d}^L$, and its contribution to the amplitude is analytically
the same as the contribution of the standard CC vertex (remember that we
have neglected quark masses). The only difference is a multiplication of
some amplitudes by mixing matrix elements. There are only two generic
variants for the case of scattering topology diagrams illustrated in
Fig.\ref{fig1}: CC  in Fig.\ref{fig1}a and NC in  Fig.\ref{fig1}b. One can
easily see that the rotation of down quarks results in only one factor
$V_{ij}$ (or $V_{ij}^{-1}$ in conjugated cases) in front of the each CC
line, while the mixing matrix elements cancel each other inside the NC line
due to the unitarity of the mixing matrix   $\sum_{k}\,V_{ik}^{-1}\,V_{kj}
= \delta_{ij}$.  One should stress  that the  summation over internal
flavours in propagators is performed  when the amplitude is evaluated.  In
the Fig.\ref{fig1} only one $W$ vertex is shown in the CC case and two
$W$ vertices in the NC case. However the statements are correct for the general
CC and NC cases. Obviously the same conclusions are valid in the case of CC
and NC diagrams of the annihilation topology as shown in Fig.\ref{fig2}b,c.

An important difference between the annihilation and scattering topologies
appears at the level of squared diagrams when the convolution with the
parton distribution functions is performed.

\vspace{0.5cm}
\noindent
{\large \bf 2. Squared diagrams and PDF convolution}
\vspace{0.2cm}

The next step is to square the matrix element and convolute with the PDF's.
Here we derive four basic rules for the technique under discussion.

\vspace{0.5cm}
\noindent
{\bf Scattering topology. 1st Rule}

Let us consider Feynman diagrams with the scattering topology. The second
{\it in}-state can be a quark, an anti-quark or a gluon. We denote the sum
of this type of Feynman diagrams as ${\cal D}_{sc}$. An example is
presented in Fig.~\ref{fig2}a for the case of a CC upper quark line and an
NC lower quark line. After squaring of diagrams of this type and
convoluting with PDF's one can write the contribution to the cross section
in the form (for the example in the Fig.~\ref{fig2}a):
\begin{eqnarray}
|{\cal D}_{sc}|^2 \;&\Longrightarrow&\; \sum_{ijkn}\,
\int dx_1 dx_2 \, f_{d_i}(x_1) \, f_{u_j}(x_2) \; |M_{ijkn}|^2\;=  \nonumber \\
  &=&\;\int dx_1 dx_2\,
 \left[ \sum_{ik}\, f_{d_i}(x_1)\,V_{ik}\,V^{-1}_{ki} \right]
 \left[ \sum_{jn}\, f_{u_j}(x_2)\,\delta_{jn}\,\delta_{nj} \right]
 |{\cal M}|^2\;,   \nonumber
\end{eqnarray}
where by $M$ and ${\cal M}$ we denote matrix elements evaluated in the
standard EW theory and in the $EW_{ud}$ model respectively. Due to the
unitarity of the mixing matrix one has $V_{ik}V_{ki}^{-1}=\delta_{ii}$, and
the first rule can be written as
\begin{equation}
     |{\cal D}_{sc}|^2 \;\Longrightarrow\; 
     \int dx_1 dx_2\; [f_d(x_1)+f_s(x_1)]\;
                      [f_{\bar u}(x_2)+f_{\bar c}(x_2)]
                   \; |{\cal M}|^2\;.
\label{rule1}
\end{equation}
This means that one can evaluate squared diagrams of the scattering
topology class with only one quark generation, but should convolute the 
corresponding gauge invariant squared matrix element with modified
structure function(s) -- a sum of down (or up) PDF's. 

\vspace{0.5cm}
\noindent
{\bf Annihilation CC topology. 2nd Rule}

In the case of the annihilation topology the CC and NC cases lead to
different rules for convolution with the PDF. We start from the CC
annihilation case which is generically shown in Fig.~\ref{fig2}b. Recall
that this variant occurs only if the quark line has an odd number of $W$
vertices. We denote the sum of this class of Feynman diagrams as ${\cal
D}_{a}^{CC}$. When one squares this sum of diagrams and convolutes with the
PDF, one gets:
\begin{eqnarray}
|{\cal D}_{a}^{CC}|^2 \;&\Longrightarrow&\; \sum_{ij}\, 
\int dx_1 dx_2 \, f_{d_i}(x_1) \, f_{\bar u_j}(x_2) \;|M_{ij}|^2  \nonumber \\
  &=& \; \int dx_1 dx_2\, \left[ \sum_{ij}\, 
    f_{d_i}(x_1)\,f_{\bar u_j}(x_2)\,V_{ij}\,V^{-1}_{ji} \right]
  |{\cal M}|^2\;.  \nonumber
\end{eqnarray}

In contrast to the scattering case here one can not use the unitarity
condition to cancel two elements of the mixing matrix because the
summations over the indices, $i$ and $j$, also include the structure functions
$f_{d_i}$ and $f_{\bar u_j}$. Thus, the second rule can be written in the
following form (see the example in Fig.~\ref{fig2}b):
\begin{eqnarray}
|{\cal D}_{a}^{CC}|^2 \;\Longrightarrow\;
 \int dx_1 dx_2 &[&f_d(x_1)\,f_{\bar u}(x_2) \, \cos^2\vartheta_c
     +f_s(x_1)\,f_{\bar c}(x_2) \, \cos^2\vartheta_c +  \label{rule2} \\
     &+& \; f_d(x_1)\,f_{\bar c}(x_2) \, \sin^2\vartheta_c
        +f_s(x_1)\,f_{\bar u}(x_2) \, \sin^2\vartheta_c
		      ] \; |{\cal M}|^2\;.  \nonumber
\end{eqnarray}
where we have explicitly expressed the mixing matrix elements
in terms of the Cabbibo mixing angle. 

One can see that the annihilation-type contribution to the squared matrix
element is convoluted with non-factorizable products of PDF's.

\vspace{0.5cm}
\noindent
{\bf Annihilation NC topology. 3d Rule}

We denote the sum of all diagrams with the annihilation NC quark line as
${\cal D}_{a}^{NC}$. The generic example is shown in Fig.~\ref{fig2}c. In
this case the quark line has an even number of $W$-boson vertices. In the
same way as above one can derive the following formula:
\begin{eqnarray}
|{\cal D}_{a}^{NC}|^2 \;&\Longrightarrow&\; \sum_{ij}\, 
\int dx_1 dx_2 \, f_{d_i}(x_1) \, f_{\bar d_j}(x_2) \;
 |M_{ij}|^2 \; =                 \nonumber \\
  &=&\;\int dx_1 dx_2\,
 \left[ \sum_{ij}\, 
    f_{d_i}(x_1)\,f_{\bar d_j}(x_2)\,\delta_{ij}\,\delta_{ji} \right]
  |{\cal M}|^2\;.     \nonumber
\end{eqnarray}

Therefore, the third rule can be written in the form (if the {\it in}-quarks are
down):
\begin{equation}
|{\cal D}_{a}^{NC}|^2 \;\Longrightarrow\;
 \int dx_1 dx_2\, [f_d(x_1)\,f_{\bar d}(x_2)
                  +f_s(x_1)\,f_{\bar s}(x_2)]
                \; |{\cal M}|^2\;.
\label{rule3}
\end{equation}
with the obvious generalization for up quarks.

\vspace{0.5cm}
\noindent
{\bf Interference of ${\cal D}_{sc}$ and ${\cal D}_{a}$ topologies}

In the general case diagrams of both, scattering and annihilation,
topologies could contribute. The interference between gauge invariant
diagram classes with these topologies is also gauge invariant and,
therefore, can be independently convoluted with the PDF. As in the
above derivations the PDF convolution for the interference of ${\cal
D}_{sc}$ diagrams with ${\cal D}_{a}^{CC}$ diagrams is given by the same
formula as for $|{\cal D}_{a}^{CC}|^2$ (2nd Rule), and for the interference
of ${\cal D}_{sc}$ with ${\cal D}_{a}^{NC}$ by the corresponding formula
for $|{\cal D}_{a}^{NC}|^2$ (3rd Rule). 

\vspace{0.5cm}
\noindent
{\bf The final state quark-antiquark line. 4th Rule}

Finally let us consider Feynman diagrams where a quark line connects two
{\it out}-states, as in Fig.~\ref{fig2}d. Here the  summation over
generation indices does not involve parton distribution functions and can,
therefore, be performed explicitly. The result is that the contribution of the
corresponding squared diagrams, evaluated in the $EW_{ud}$ model, should be
multiplied by two for both the CC and NC cases. Indeed, according to
Fig.~\ref{fig2}d each of these summations give in the squared diagram the
factors: $\sum_{ij}\,\delta_{ij}\,\delta_{ji}=2$ in the NC and 
$\sum_{ij}\,V_{ij}\,V^{-1}_{ji}=\sum_{ij}\,\delta_{ij}=2$ in the CC cases.

Note that the 4th Rule is valid not only in cases where the  quark loop in
squared diagrams connects {\it out}-state(s) and never passes through {\it
in}-state(s), but also for each quark loop in next-to-leading corrections.

Of course, the 4th Rule is valid only under the  assumption that the
fragmentation of the four light quarks and antiquarks leads to
indistinguishable jets. If one includes nontrivial fragmentation
functions, e.g. for a $c$-quark, all the above rules have to be modified.
We do not present here the corresponding formulas which however could be
easily derived.

\vspace{0.5cm}
\noindent
{\large \bf 3. Test: $W+2jets$ production at LHC}
\vspace{0.2cm}

In this section we illustrate the proposed technique  with the example of
$W^+ +2jets$ production at LHC. Here a total of 180  subprocesses
contribute in the standard technique when all four light quarks contribute
separately and the CKM matrix is present in $W$-boson vertices. With the new
technique only 21 subprocesses need to be evaluated.

We shall not discuss this example in full detail but present the results of
a numerical test for the $u \bar{d}\to d \bar{d} W^+$ subprocess with
permutations of quarks within pairs ($u$,$c$) and ($d$,$s$). Note that in
this example three rules are used: 1st, 2nd and 4th. In the standard
technique 12 subprocesses are involved and the corresponding contributions
to the cross section are collected in Tab.~\ref{tab1} for two values of the
kinematical cut on the transverse momenta of the final partons and the $W$
boson, $p^{jet}_T>p^0_T$ and $p^W_T>p^0_T$ with $p^0_T=20$ or 200 GeV. The
cross sections were calculated in the Standard Model with the averaged
values for CKM matrix elements and quark masses taken from the Particle
Data Group \cite{PDG}. For calculations the CompHEP program \cite{CompHEP}
has been used, and the accumulated MC error in all cases was less than
0.6\%. We neglect the contributions of subleading diagrams with electroweak
boson propagators, calculating the cross section to leading
$\alpha\alpha_s^2$ order.

Using the new method one should evaluates only one subprocess in the one
quark doublet $EW_{ud}$ model, $q_u \bar{q_d}\to q_d \bar{q_d} W^+$. The
corresponding Feynman diagrams are shown in Fig.~\ref{fig3}. This
subprocess is of the mixed type where two gauge invariant classes of
diagrams: the annihilation CC topology (Fig.~\ref{fig3}a) and the
scattering topology (Fig.~\ref{fig3}b), contribute. The 1st Rule is used
for the squared scattering topology contribution. The 2nd Rule is used for
the squared annihilation CC topology contribution with a multiplication by
the factor 2 according to the 4th Rule. Finally, the 2nd Rule is used to
evaluate the interference between the two gauge invariant classes of
Feynman diagrams. We have calculated all three contributions to the cross
section of the subprocess  $q_u \bar{q_d}\to q_d\bar{q_d} W^+$ using
CompHEP code in which the $EW_{ud}$ model and Rules 1-3 have been
implemented (see the Appendix for details). The results (see
Tab.~\ref{tab2}) for the total rate, $\sigma(p_T>20\mbox{GeV})=112.95$ pb
and $\sigma(p_T>200\mbox{GeV})=0.30213$ pb, are in an agreement with the
"standard" calculations of Tab.~\ref{tab1} within the  statistical error of
less than 0.6\%.

\vspace{0.5cm}
\noindent
{\large \bf Conclusions}
\vspace{0.2cm}

We have shown that hard collision processes at hadron colliders can be
evaluated in an economical way, greatly reducing the number of contributing
subprocesses. The proposed computational technique can be applied  only if
the quark masses of the  first two generations and mixing with the 3rd
generation can be neglected. These assumptions are valid for most
applications at the Tevatron and LHC. In the proposed technique the
Standard Model with a single generation of up and down quarks is used and a
squared matrix element is evaluated without involving elements of the mixing
matrix. The resulting matrix squared matrix element is convoluted  with
modified parton distribution functions according to formulas  given above
as Rules 1-4. Each of these Rules corresponds to a gauge invariant class of
squared diagrams.

In Rule 1 the squared matrix element is convoluted with $f_u(x)+f_c(x)$
or $f_d(x)+f_s(x)$ for an {\it in}-state  of up-type or down-type
quarks respectively However, in the cases of Rules 2 and 3 the squared
matrix element is convoluted over Bjorken variables $x_1$ and $x_2$ with a
{\it non-factorizable} function. When applying the 2nd Rule the function
$$ 
   [f_d(x_1)\,f_{\bar u}(x_2)
  +f_s(x_1)\,f_{\bar c}(x_2)] \, \cos^2\vartheta_c
  +[f_d(x_1)\,f_{\bar c}(x_2) 
  +f_s(x_1)\,f_{\bar u}(x_2)] \, \sin^2\vartheta_c \;,
$$
or a similar function where the substitution  $ d \rightarrow \bar{d}$
and $ \bar{u} \rightarrow u$ are made, is used.  In the case of the 3rd
Rule the convolution is with  either
$$ f_u(x_1)\,f_{\bar u}(x_2)+f_c(x_1)\,f_{\bar c}(x_2)\; \quad or,
$$
$$ f_d(x_1)\,f_{\bar d}(x_2)+f_s(x_1)\,f_{\bar s}(x_2)\;,
$$
depending on which quark-antiquark pair occurs in the {\it in}-state.

\vspace{0.5cm}

The authors are indebted to A.E.~Pukhov, Th.~Ohl and B.~Straub for useful
discussions. This work was partially supported by the CERN-INTAS 99-377,
RFBR-DFG 99-02-04011 and RFBR 00-01-00704 grants, by the Russian Ministry of
Science and Technology, by St.Petersburg Grant Center and by the program
"Universities of Russia" (grant 990588).  

\def\ijmp#1#2#3{{\it Int. Jour. Mod. Phys. }{\bf #1~} (19#2) #3}
\def\pl#1#2#3{{\it Phys. Lett. }{\bf B#1~} (19#2) #3}
\def\zp#1#2#3{{\it Z. Phys. }{\bf C#1~} (19#2) #3}
\def\prl#1#2#3{{\it Phys. Rev. Lett. }{\bf #1~} (19#2) #3}
\def\rmp#1#2#3{{\it Rev. Mod. Phys. }{\bf #1~} (19#2) #3}
\def\prep#1#2#3{{\it Phys. Rep. }{\bf #1~} (19#2) #3}
\def\pr#1#2#3{{\it Phys. Rev. }{\bf D#1~} (19#2) #3}
\def\np#1#2#3{{\it Nucl. Phys. }{\bf B#1~} (19#2) #3}
\def\mpl#1#2#3{{\it Mod. Phys. Lett. }{\bf #1~} (19#2) #3}
\def\arnps#1#2#3{{\it Annu. Rev. Nucl. Part. Sci. }{\bf #1~} (19#2) #3}
\def\sjnp#1#2#3{{\it Sov. J. Nucl. Phys. }{\bf #1~} (19#2) #3}
\def\jetp#1#2#3{{\it JETP Lett. }{\bf #1~} (19#2) #3}
\def\app#1#2#3{{\it Acta Phys. Polon. }{\bf #1~} (19#2) #3}
\def\rnc#1#2#3{{\it Riv. Nuovo Cim. }{\bf #1~} (19#2) #3}
\def\ap#1#2#3{{\it Ann. Phys. }{\bf #1~} (19#2) #3}
\def\ptp#1#2#3{{\it Prog. Theor. Phys. }{\bf #1~} (19#2) #3}
\def\spu#1#2#3{{\it Sov. Phys. Usp.}{\bf #1~} (19#2) #3}
\def\apj#1#2#3{{\it Ap. J.}{\bf #1~} (19#2) #3}
\def\epj#1#2#3{{\it Eur.\ Phys.\ J. }{\bf C#1~} (19#2) #3}
\def\pu#1#2#3{{\it Phys.-Usp. }{\bf #1~} (19#2) #3}

\vspace{1.5cm}
\noindent
{\Large \bf Appendix}
\vspace{0.5cm}

The Rules derived in this letter have been implemented in the CompHEP
code v.33. New model, referred to above as $EW_{ud}$, was created
on the base of the SM where only one quark generation, 
up and down quarks denoted as $q_u$ and $q_d$
was kept without any CKM
matrix elements.
The masses of these generalized quarks were set to
zero. Then, the option for numerical convolution of squared diagrams with
parton
distribution functions was modified in accordance with Rules 1-3.
The code of this version of
CompHEP is available from the Web \cite{qQ}.

In the most general case the user has to subdivide the whole set of squared
diagrams into two parts: 1) $|{\cal D}_{sc}|^2$, and 2) $|{\cal D}_{a}|^2$
plus the interference diagrams $2 Re({\cal D}_{sc}\cdot {\cal D}_{a}^*)$.
Each of these parts should be calculated separately. In particular,
for each part the
user has to set the variable "PDFfactor" in the menu option "User menu" as
follows: $PDFfactor=1$ for $|{\cal D}_{sc}|^2$ (referred to in CompHEP as the
"t-channel" case), and $PDFfactor=0$ for $|{\cal D}_{a}|^2 + 2 Re({\cal
D}_{sc}\cdot {\cal D}_{a}^*)$ (referred to in CompHEP as the "s-channel" case).
 The
program automatically recognizes which Rule, 2nd or 3rd, should be used in
the latter case.

Rule 4 (multiplication by a factor 2) has to be applied by hand if the
corresponding quark line is presented in a squared diagram.

One should note that this is not a completely automatic realization of
Rules 1-4. This is planned for future CompHEP development.

\clearpage
\noindent
{\Large \bf Figures}

\clearpage
\vspace*{0.4cm}

\vspace*{1.7cm}

\clearpage
\noindent
{\Large \bf Tables}

\begin{table}[hb]
\begin{center}
\begin{tabular}{|c|ll|}
\hline
 Subprocess        & \multicolumn{2}{c|}{cross section, pb} \\
\cline{2-3}
                             & $p^0_T=20$ GeV &  $p^0_T=200$ GeV  \\   
\hline
$u\bar d \to s \bar s W^+$   & 9.368    &    0.01746  \\
$u\bar d \to \bar d s W^+$   & 2.315    &    0.007807  \\
$u\bar d \to d \bar d W^+$   & 55.84    &    0.1733  \\
$u\bar s \to s \bar s W^+$   & 2.113    &    0.005301  \\
$u\bar s \to d \bar s W^+$   & 32.67    &    0.09074  \\
$u\bar s \to d \bar d W^+$   & 0.3748   &    0.0005516  \\
$c\bar d \to s \bar s W^+$   & 0.0739   &    0.00003557  \\
$c\bar d \to \bar d s W^+$   & 4.601    &    0.004845  \\
$c\bar d \to d \bar d W^+$   & 0.3206   &    0.0002915  \\
$c\bar s \to s \bar s W^+$   & 4.093    &    0.003090  \\
$c\bar s \to d \bar s W^+$   & 0.1528   &    0.0001359  \\
$c\bar s \to d \bar d W^+$   & 0.9842   &    0.0003853  \\
\hline
  In total                   & 112.90   &   0.30391   \\
\hline
\end{tabular}
\end{center}
\caption{Contributions, in pb, of different $q\bar{q'}$ channels to the
$W^+ +2jets$ production cross section at LHC. 
In addition to cuts on transverse
momenta,
$p^{jet}_T>p^0_T$ and $p^W_T>p^0_T$, the following cuts were applied: on
pseudorapidity $|\eta_{jet}|<5$ and on the jet separation
in the $\eta-\varphi$ plane $R(jet,jet')>0.5$. The statistical accuracy of the
MC calculations is better than 0.6\%.
\label{tab1} }
\end{table}

\begin{table}[hb]
\begin{center}
\begin{tabular}{|c|ll|}
\hline
Classes of squared diagrams, &\multicolumn{2}{c|}{contribution to cross section,
pb}\\
\cline{2-3}
 used Rules                       & $p^0_T=20$ GeV&  $p^0_T=200$ GeV   \\
\hline
$|{\cal D}_{sc}|^2$,  1st Rule              & 89.76   &      0.2621    \\
$|{cal D}_{a}^{CC}|^2$, 2nd + 4th Rules    & 21.62   &      0.03692   \\
$2Re({\cal D}_{a}^{CC}\cdot {\cal D}_{sc}^*)$, 2nd Rule & 1.549 & 0.003111  \\
\hline
  In total                        & 112.93  &      0.30213  \\
\hline
\end{tabular}
\end{center}
\caption{Contribution, in pb, of the subprocess 
$q_u \bar{q_d}\to q_d \bar{q_d} W^+$ to the
$W^+ +2jets$ production cross section at LHC, evaluated with the help of
Rules 1,2 and 4 using the of $EW_{ud}$ model. The same cuts were applied as in 
{\protect Tab.~\ref{fig1}}. The accuracy of the MC calculations is better than 
0.6\%.
\label{tab2} }
\end{table}

\end{document}